# Quantum coherence across bosonic superconductor-anomalous metal-insulator transitions

Chao Yang[1,*], Yi Liu[2,3,*], Yang Wang[1], Liu Feng[1], Qianmei He[1], Jian Sun[2,3], Yue Tang[2,3], Chunchun Wu[1], Jie Xiong[1†], Wanli Zhang[1], Xi lin[2,3], Hong Yao[3,4,5], Haiwen Liu[6], Gustavo Fernandes[7], Jimmy Xu[7,8], James M. Valles Jr.[8†], Jian Wang[2,3,5,9,10†], and Yanrong Li[1]

[1]State Key Laboratory of Electronic Thin Films and Integrated Devices, University of Electronic Science and Technology of China, Chengdu 610054, China.

[2]International Center for Quantum Materials, School of Physics, Peking University, Beijing 100871, China.

[3]Collaborative Innovation Center of Quantum Matter, Beijing 100871, China.

[4]Institute for Advanced Study, Tsinghua University, Beijing 100084, China.

[5]State Key Laboratory of Low Dimensional Quantum Physics, Tsinghua University, Beijing 100084, China.

[6]Center for Advanced Quantum Studies, Department of Physics, Beijing Normal University, Beijing 100875, China.

[7]School of Engineering, Brown University, 182 Hope Street, Providence, RI02912, USA.

[8]Department of Physics, Brown University, 182 Hope Street, Providence, RI 02912, USA.

[9]CAS Center for Excellence in Topological Quantum Computation, University of Chinese Academy of Sciences, Beijing 100190, China.

[10] Beijing Academy of Quantum Information Sciences, West Bld. #3, No. 10 Xibeiwang East Rd., Haidian District, Beijing 100193, China

*These authors contributed equally to this work.

†Corresponding author. E-mail: jianwangphysics@pku.edu.cn (J.W.); james_valles_jr@brown.edu (J.M.V.J); jiexiong@uestc.edu.cn (J.X.)

**Abstract**:

After decades of explorations, suffering from low critical temperature and subtle nature, whether a metallic ground state exists in a two-dimensional system beyond Anderson localization is still a mystery. Supremely, phase coherence could be the key that unlocks its intriguing nature. This work reveals how quantum phase coherence evolves across bosonic superconductor-metal-insulator transitions via magneto-conductance quantum oscillations in high-Tc superconducting films. A robust intervening anomalous metallic state characterized by both resistance and oscillation amplitude saturations in the low temperature regime is detected. By contrast, with decreasing temperature the oscillation amplitude monotonically grows on the superconducting side, but decreases at low temperatures on the insulating side. It suggests that the saturation of phase coherence plays a prominent role in the formation of this anomalous metallic state.

**Main Text:**

In 1979, the scaling theory of localization predicted that metallic states are absent at zero temperature in two-dimensional (2D) disordered systems due to quantum interference (*1*). The phase coherence length as a fundamental property will diverge as electron waves are exponentially localized (*1-2*). Consistently when electrons form Cooper pairs in 2D disordered system, the ground states should be superconductors or insulators. Nevertheless, approaching zero temperature, an anomalous metallic ground state has been experimentally observed (*3-15*), which shows a residual resistance far below the quantum resistance at low temperatures. The existence and origin of this unexpected metallic state gives rise to intense debates (*16-24*).

The resistance saturation as the characteristic of anomalous metallic ground state in 2D superconductors (*3-15*) has been experimentally observed in ultralow temperature regime. However, extrinsic effects such as noise and carrier overheating may also disrupt the phase coherence, which raises the difficulty of clarifying the origin of 2D metal-like state at zero temperature (*25*). To date, the nature of the anomalous metallic state is still mysterious and the conclusive evidence is highly desired. The phase coherence could be a key to understand 2D metallic state. If the phase coherence length of electrons saturates at low temperatures, due to the weakening of constructive quantum interference, the diffusion should not be absolutely prohibited resulting in a possible metallic state in a low dimensional electron system (*26-28*). Thus, it is crucial to investigate the phase coherence of Cooper pairs in anomalous metallic state of 2D superconductors.

Here, we present a systematic study of Cooper pair phase coherence through the superconductor-anomalous metal phase transition followed by the anomalous metal-insulator phase transition in high-Tc superconductor (HTS) films patterned with array of holes. The relatively high energy scale of HTS and low temperature measurements make the observations reliable. The wide resistance saturation over one decade of temperature variation, zero Hall resistance and the giant magneto-resistance are simultaneously detected as characteristics of an anomalous metallic state. Interestingly, magneto-conductance quantum oscillations with a period of h/2e are detected in superconducting, anomalous metal and insulating states, providing a direct observable of Cooper pair coherence through the phase transitions.

By reactive ion etching (RIE), we etched a high quality 12 nm thick YBCO thin film through a contact mask with a hexagonal array of holes (see Fig. S1). Each mask is an anodized aluminum oxide (AAO) membrane formed on and then separated from an aluminum base (see Fig. 1A). The approach offers the advantages of minimal molecular residuals and processing contamination (*29-31*).

The RIE transfers the pattern, around 70 nm diameter holes with center-to-center hole spacing of ~103 nm, onto the HTS film (Fig. 1B). In the process, the ion bombardment impacts the sidewalls of the resultant pores to create a less crystalline and strained YBCO shell, which becomes thicker with increasing etching time (*31, 32*). This was confirmed in a detailed high-resolution transmission-electron microscopy (HRTEM) analysis, which shows high density of structural defects around the nanopatterned holes (*33*). We deduce from the transport data presented below that this process forms superconducting islands at the nodes of the hexagonal pore lattice connected to three links. The islands shrink and the links become more resistive with increasing etching time (Fig. 1C). The measured $R_S(T)$ curves of the nanopatterned YBCO thin films exhibit quantum phase transition (QPT) behaviors, as shown in Fig. 1D. The YBCO thin film subjected to a minimal RIE time remains superconducting (see a representative film SC exhibited in Fig. 1D). The superconducting onset critical temperature $T_c^{onset}$ is defined as the temperature at which the sheet resistance $R_S(T)$ curve deviates from the linear extrapolation of the normal state (Fig. 1D & Fig. S2) and the normal state sheet resistance is defined as $R_N = R_S(T_c^{onset})$. With increasing etching time, the superconducting transition region becomes wider and the zero resistance state fades away. We define this state as anomalous metal (AM) state as there is a plateau of residual resistance at lower temperatures. A representative AM film is shown in Fig. 1D. For films with the normal resistance larger than the critical value $R_S \approx 13$ k$\Omega$, $R_S(T)$ presents a superconducting transition $(dR_S/dT > 0)$ below $T_c^{onset}$ followed by an insulating transition $(dR_S/dT < 0)$ at lower temperatures. We define this as a transitional state (TS). A typical TS curve is marked in Fig. 1D. The samples with more etching exhibit a more pronounced insulating upturn emerging at higher temperatures. With further etching, $R_S(T)$ increases monotonically $(dR_S/dT < 0)$ to infinite at zero temperature indicating an insulating property over the whole temperature range. An example labeled as INS (insulating state) is shown in Fig. 1D. More details of electrical transport of all films can be found in Fig. S2.

Figure 2 shows the characteristics of the anomalous metal state in HTS films with an array of nanopores. The Arrhenius plot of the anomalous metallic state films and superconducting films are presented in Fig. 2A. For superconducting films, the resistance below $T_c^{onset}$ drops to zero within the measurement resolution. For metallic state films, in the low temperature regime, the resistance drops and saturates to a finite value which can be up to five orders of magnitude smaller than normal state resistance. In the smallest saturating resistance film, the saturation of resistance starts at around 5 K, which is much higher than that of conventional superconductors (*4-8, 10-15*). To exclude external noise effects, we measured metallic state films using well-filtered electrical leads in a dilution refrigerator cryostat down to 50 mK as shown in Fig. 2B & Fig. S3. The I-V curves are linear below 100nA indicating the measurements are in the ohmic regime, as shown in inset of Fig. 2B. The resistance saturation remains almost the same with or without the resistor-capacitor (RC) filters, demonstrating that the metallic state is intrinsic.

To further explore the nature of the anomalous metallic state, the Hall resistance ($R_{xy}$) and longitudinal resistance ($R_{xy}$) were measured simultaneously. Below $T_c^{onset}$, the slopes of $R_{xy}(\mu_0 H)$ curves are largely suppressed and drop to zero with decreasing temperature for a representative metallic state film (Fig. 2C inset). Figure 2C shows the temperature dependence of $R_{xy}$, which becomes zero within the measurement resolution at a relatively high temperature ~50 K, while $R_{xx}$ is still finite. The absence of Hall resistance at low temperatures reveals the particle–hole symmetry in the anomalous metal (also see Fig. S4). Furthermore, a giant positive longitudinal magneto-resistance is simultaneously observed as another important feature of this anomalous metallic state *(13)* (Fig. 2D).

Below $T_c^{onset}$, the perpendicular magneto-conductance of all the YBCO nanopore thin films in Fig. 1D oscillates at low fields as shown in Fig. 3A-C for three representative films. We present the data as the negative change of magneto-conductance $-\Delta G = G_0 - G(B)$, where $G_0$ is the conductance at zero magnetic field. The magneto-conductance of all films is symmetric and the magneto-conductance oscillations appear on a monotonically rising background (full data shown in Fig. S5). Up to four magneto-conductance oscillations are detected with a constant period of $H_0$~2.25 kOe. This period is consistent with one superconducting flux quantum $\Phi_0 = h/2e$ threading an area of one unit-cell of the hexagonal pattern (around 9200 nm$^2$), where *h* is Planck's constant and *e* is the electron charge. The appearance of the charge 2*e* quantum oscillations just below $T_c^{onset}$ demonstrate that Cooper pairs participate in the transport in all the states through the superconductor-metal-insulator transitions.

To highlight the temperature dependence of the oscillations, we define the magneto-conductance oscillation amplitude as $G_{osc} = |G(1/2H_0) - G_0|$ after subtracting the rising background. Figure 3D shows the temperature dependence of $G_{osc}$ on a logarithm scale for the films exhibiting superconducting, anomalous metallic and insulating states, respectively. For the films of superconducting state, the oscillation amplitude monotonically grows to a giant value beyond our measurement limit with decreasing temperature. In contrast, the oscillation amplitude saturates in the anomalous metallic region below 5 K (Fig. 3D), reminiscent of the residual resistance plateau of anomalous metallic state at low temperatures (Fig. 2A). For insulating films (i.e. INS films), with decreasing temperature $G_{osc}$ peaks and then drops. Interestingly, with increasing disorder, $G_{osc}$ varies by at least nine orders of magnitude from the superconducting films to most insulating films in the zero temperature limit, while the normal resistance only changes two orders of magnitude. It demonstrates that the Cooper pair coherence over a wide range can be controlled by tuning the link resistance of nanopatterned YBCO films.

The electrical transport and the magneto-conductance oscillations in YBCO nanopore thin films can be qualitatively understood by modeling them as resistively shunted Josephson junction arrays. With increasing etching time, the link resistances grow and the superconducting islands shrink *(34),* which drives the system through a superconductor-anomalous metal-insulator transition. For all films, $T_c^{onset}$ varies by no more than 15% from the superconducting film to the insulating film, which indicates that the strength of local superconducting pairing remains stable through the QPT while Cooper pair phase coherence can be tuned by increasing link resistance.

The magneto-conductance oscillations originate from Cooper pair quantum interference effects in the multiply connected geometry imposed by the array structure *(35,36)*. These effects enter through the Josephson coupling energy between nodes of the array as given by $E_J = -J cos(\phi_i - \phi_j - A_{ij})$, where $\phi_i - \phi_j$ is the gauge invariant phase difference of the order parameter between islands *i* and *j*, and $A_{ij} = h/2e \int_j^i \boldsymbol{A} \cdot \boldsymbol{dl}$ is the line integral of the vector potential between neighboring nodes.

*J* is proportional to the order parameter amplitude on the islands. Josephson coupling energy $E_J$ oscillates with increasing magnetic field due to the requirement that the line integral of the phase around a closed loop is a multiple of 2π. The dimensions of the array unit dictate the period of the oscillations. The magneto-conductance oscillates on the superconducting side because $E_J$ is the barrier to thermally activated vortex motion *(36)*. The magneto-conductance oscillates on the insulating and anomalous metal side because $E_J$ is proportional to the Cooper pair tunneling rate.

We now discuss the evolution of the temperature dependence of the oscillations across the QPT within this array model. For all the YBCO films, the oscillations appear just below $T_c^{onset}$ indicating that local phase coherence develops as soon as the amplitude of the order parameter forms on the islands. At lower temperatures, the oscillation amplitude is approximately proportional to the phase coherence as the growth of order parameter on the islands ceases. For superconducting films, the oscillation amplitude monotonically grows beyond our measurement range with decreasing temperature, indicating the phase coherence length diverges in the zero temperature limit. By contrast, the oscillation amplitude of anomalous metallic state films saturates below around 5 K after a severe growth. The saturating temperature dependence of the oscillation amplitude is consistent with that of the resistance, indicating that phase coherence saturation should be crucial to the formation of the anomalous metallic state. For the insulating films, as the link resistance rises beyond $R_Q = h/4e^2$ with decreasing temperature, the interisland capacitance and self-capacitances of the islands can produce a Coulomb blockade to Cooper pair transport *(37)*. Thus $G_{osc}$ and the phase coherence length decrease as the Cooper pairs become localized. In order to quantify phase coherence through the QPT, we estimate the phase coherence length $L_\phi$ by approximately using the formula $G_{\text{osc}} = \frac{4e^2}{h} (\frac{L_\phi}{\pi r})^{1.5} \exp(-\frac{\pi r}{L_\phi})$ for quasi-particle quantum interference *(28,38)*, where r is half of center-to-center hole spacing ~ 50 nm, *h* is Planck's constant and *e* is the electron charge. It is noteworthy to mention that the above formula can only be applied to insulating or some metallic states films with relatively larger residual resistance when $G_{\text{osc}}$ is smaller or comparable to quantum conductance. The temperature dependence of $L_\phi$ are presented in Fig. 3E, indicating the saturation of quantum coherence for anomalous metallic states in the low temperature regime. It is noted that the magneto-conductance oscillation amplitude observed in the anomalous metallic states with small residual resistance is giant and much larger than the quantum conductance, which is hard to be simply explained by quasi-particle quantum interference, leaving an open question for further investigations.

The phase diagrams of resistance and differential resistance as a function of temperature and normal resistance $R_N$ in Fig. 4 provide a summary of the results. We use $R_N$ as the tuning parameter since it is directly related to the etching time (see Fig. S6) and increases with increasing disorder in the nanopatterned films *(34)*. The normal state and other states are separated by $T_c^{onset}$. At the boundary from $dR_s/dT > 0$ to $dR_s/dT = 0$, the array resistance drops below $T_c^{onset}$ to zero (SC) or a plateau of residual resistance (AM) at the lowest temperatures as shown in the bottom-left part of the phase diagram (Fig. 4). The crossover between the AM and resistance drop region is separated by a short dash line (ill defined). Below $T_c^{onset}$, the onset of insulating behavior is visible in the lower right region where $dR_s/dT$ is negative. We label this region as the Cooper pair insulator (CPI). Approaching zero temperature, with increasing $R_N$, there is a regime where $dR_s/dT$ changes from zero to a negative value which can be interpreted as the critical regime of a quantum phase transition. The phase diagram shows the evolution of anomalous metal state to a Cooper pair insulator state separated by a QCP around $R_N$~13 kΩ.

To date, the nature of the anomalous metallic state in 2D system is still mysterious. Candidate theory includes dissipating bath coupled to environment *(16-18)*, Bose metal *(19)*, order parameter fluctuation *(20-23)* and composite Fermi liquid phase *(24)*. In our HTS system, this anomalous metallic state is clearly a bosonic state demonstrated by h/2e quantum oscillations. As for the fermionic system, the phase coherence saturation is argued to be the key to understand low-dimensional metal *(26-28, 39)*, which might be universal for both fermionic and bosonic system. However, the very low phase coherence saturation temperature in fermionic system creates challenges to explore the underlying mechanism. An anomalous metal state appears in a d wave superconductor at higher temperatures with zero Hall resistance indicates that the low energy excitation spectrum plays a secondary role in its formation and the model on Cooper pair phase coherence may be the right direction to reveal the nature of this anomalous metallic state. Thus, our work paves the way for investigations on the anomalous state of matter by using finite frequency probes, like terahertz spectroscopy that are more difficult to apply at dilution refrigerator temperatures. The ability to control and reveal the quantum coherence and relatively high critical temperature in our nanopatterned HTS system will shed light on the anomalous metallic state and quantum phase transitions in low-dimensional systems.

**References and Notes:**

1. E. Abrahams, P. W. Anderson, D. C. Licciardello, T. V. Ramakrishnan, *Phys. Rev. Lett.* **42**, 673-676 (1979).

2. B. L. Altshuler, A. G. Aronov, *in Electron–Electron Interactions in Disordered Systems* A. L. & Pollak. M, Eds. (North-Holland, 1985).

3. H. M. Jaeger, D. B. Haviland, B. G. Orr, A. M. Goldman, *Phys. Rev. B* **40**, 182–196 (1989).

4. E. Shimshoni, A. Auerbach, A. Kapitulnik, *Phys. Rev. Lett.* **80**, 3352–3355 (1998).

5. D. Ephron, A. Yazdani, A. Kapitulnik, M. R. Beasley, *Phys. Rev. Lett*. **76**, 1529–1532 (1996).

6. B. Spivak, A. Zyuzin, M. Hruska, *Phys. Rev. B* **64**, 132502 (2001).

7. C. Christiansen, L. M. Hernandez, A. M. Goldman, *Phys. Rev. Lett.* **88**. 037004 (2002).

8. S. Eley, S. Gopalakrishnan, P. M. Goldbart, N. Mason, *Nat. Phys.* **8**, 59–62 (2012).

9. J. Garcia-Barriocanal, *et al., Phys. Rev. B* **87**, 024509 (2013).

10. Z. Han, *et al., Nat. Phys*. **10**, 380–386 (2014).

11. Y. Saito, Y. Kasahara, J. Ye, Y. Iwasa, T. Nojima, *Science* **350**, 409–413 (2015).

12. Nicholas P. Breznay, Aharon. Kapitulnik, *Sci Adv*. **3**, e1700612 (2017).

13. A. Kapitulnik, S. A. Kivelson, B. Spivak, *Preprint at* https://arxiv.org/abs/1712.07215 (2017).

14. C. G. L. Bøttcher, *et al.*, *Nat. Phys*. **14**, 1138-1144 (2018).

15. Z. Chen, *et al.*, *Nat. Com*. **9**, 4008 (2018).

16. E. Shimshoni, A. Auerbach, A. Kapitulnik, *Phys. Rev. Lett*. **80**, 3352–3355 (1998).

17. A. Kapitulnik, N. Mason, S. A. Kivelson, S. Chakravarty, *Phys. Rev. B* **63**, 125322 (2001).

18. V. M. Galitski, G. Refael, M. P. A. Fisher, T. Senthil, *Phys. Rev. Lett.* **95**, 077002 (2005).

19. P. Phillips, D. Dalidovich, *Science* **302**, 243–247 (2003).

20. B. Spivak, A. Zyuzin, M. Hruska, *Phys. Rev. B* **64**, 132502 (2001).

21. P. Goswami, S. Chakravarty, *Phys. Rev. B* **73**, 094516 (2006).

22. A. Punnoose, A. M. Finkel'stein, *Science* **310**, 289–291 (2005).

23. B. Spivak, P. Oreto, S. A. Kivelson, *Phys. Rev. B* **77**, 214523 (2008).

24. M. Mulligan, S. Raghu, *Phys. Rev. B* **93**, 205116 (2016).

25. I. Tamir, *et al., arXiv preprint arXiv:* 1804.04648 (2018).

26. P. Mohanty, E. M. Q. Jariwala, R. A. Webb, *Phys. Rev. Lett*. **78**, 3366-3369 (1997).

27. J. J. Lin, J. P. Bird, *J. Phys. Condens. Matter* **14**, R501 (2002).

28. F. Pierre, *et al.*, *Phys. Rev. B* **68**, 085413 (2003).

29. J. Liang, H. Chik, A. Yin, J. Xu, *J. Appl. Phys.* **91**, 2544-2546 (2002).

30. H. Chik, J. M. Xu, *Mater. Sci. Eng. R*. **43**, 103-138 (2004).

31. S. G. Cloutier, P. A. Kossyrev, J. Xu, *Nat. Mater.* **4**, 887-891 (2005).

32. S. A. Cybart, *et al.*, *Nat. Nanotech*. **10**, 598-602 (2015).

33. S. G. Cloutier, C. H. Hsu, P. A. Kossyrev, J. Xu, *Adv. Mater.* **18**, 841-844 (2006).

34. J. M. Valles, *et al.*, *Phys. Rev. B* **39**, 11599 (1989).

35. M. D. Stewart, A. Yin, J. M. Xu, J. M. Valles, *Science* **318**, 1273-1275 (2007).

36. I. Sochnikov, A. Shaulov, Y. Yeshurun, G. Logvenov, I. Božović, *Nat. Nanotech.* **5**, 516-519 (2010).

37. Y. Takahide, H. Miyazaki, A. Kanda, Y. Ootuka, J. Phys. Soc. Jpn. **72**, 96-99 (2003).

38. E. Akkermans, G. Montambaux *Mesoscopic Physics of Electrons and Photons* (Cambridge Univ. Press, Cambridge, 2007).

39. S. Washburn, R.A. Webb, *Adv. In. Phys.* **35**, 375-422 (1986)

**Acknowledgements:** We thank S. A. Kivelson, J. M. Kosterlitz, D.Oller, D.He, J. H. Kim, D. J. Lee, K. Liu and J. Qi for the fruitful discussions. This work was supported by the National Natural Science Foundation of China (Grant No.11888101), the National Basic Research Program of China (Grant No.2015CB358600, No. 2018YFA0305604, No.2017YFA0303300, No. 2016YFA0301001, and No. 2017YFA0304600), the National Natural Science Foundation of China (Grant No.51722204, No.11774008, No.91421110, No. 11474175 and No. 11674028), Fundamental Research Funds for the Central Universities (ZYGX2016Z004), the Strategic Priority Research Program of Chinese Academy of Sciences (Grant No. XDB28000000), the Open Research Fund Program of the State Key Laboratory of Low-Dimensional Quantum Physics, Tsinghua University (Grant No. KTS01703), Beijing Natural Science Foundation (Z180010) and NSF DMR-1408743, CMMI-1530547, and ARO W911NF-14-2-0075

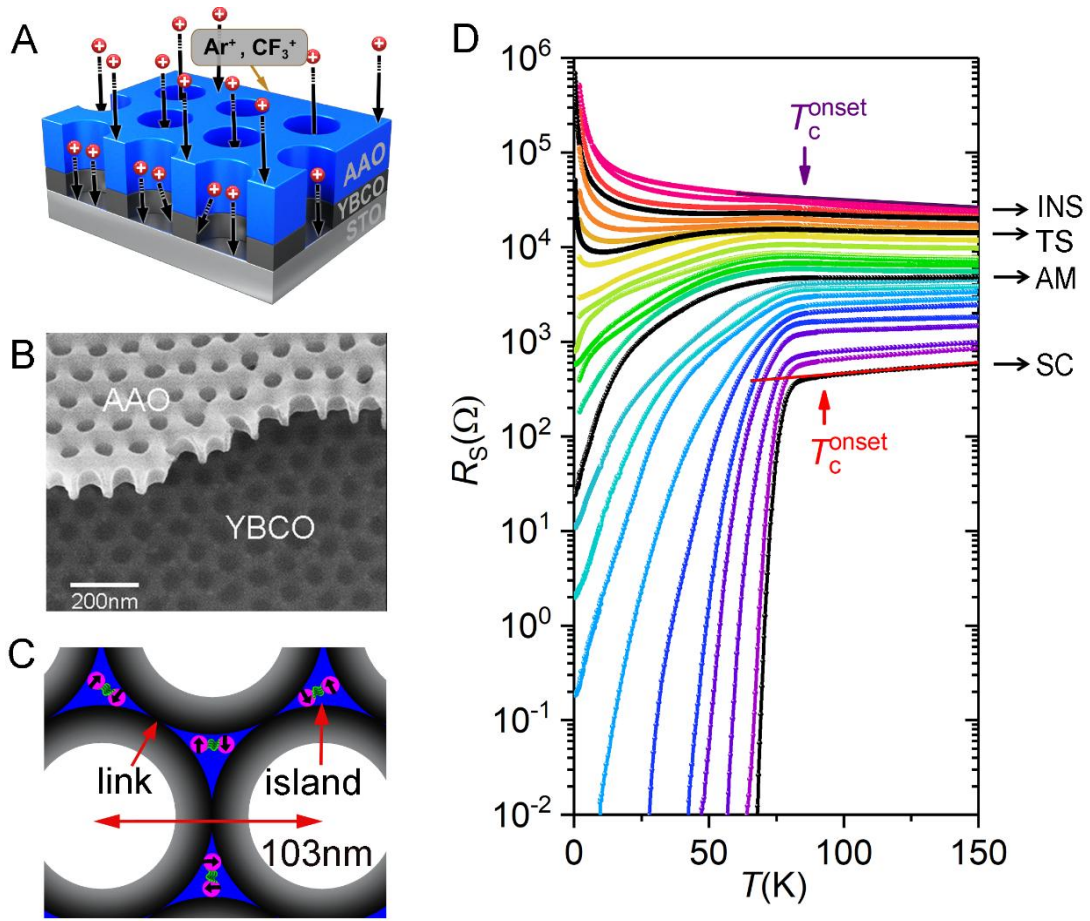

**Fig. 1. Superconductor-anomalous metal-insulator transition in nanopore modulated YBCO thin films.** (**A**) Artistic representation of the fabricating process of nanopore modulated YBCO thin films by reactive ion etching patterned by AAO. (see Methods for details) (**B**) SEM image of etched nanopore YBCO films with remained AAO pattern. (**C**) Artistic representation of etched nanopore YBCO films: the gray halos represent the weakened part given by the ion bombardment and etching, blue triangle regions represent the SC islands. Cooper pairs are exhibited by purple sphere pairs. (**D**) Temperature dependence of resistance for the etched nanopore YBCO films with different etching time. The nanopore YBCO films undergo QPT with different normal sheet resistance by tuning the etching time. Four representative nanopore YBCO films marked as superconducting (SC) state film, anomalous metallic (AM) state film, transitional (TS) state film and insulating (INS) state film are shown in black curves.

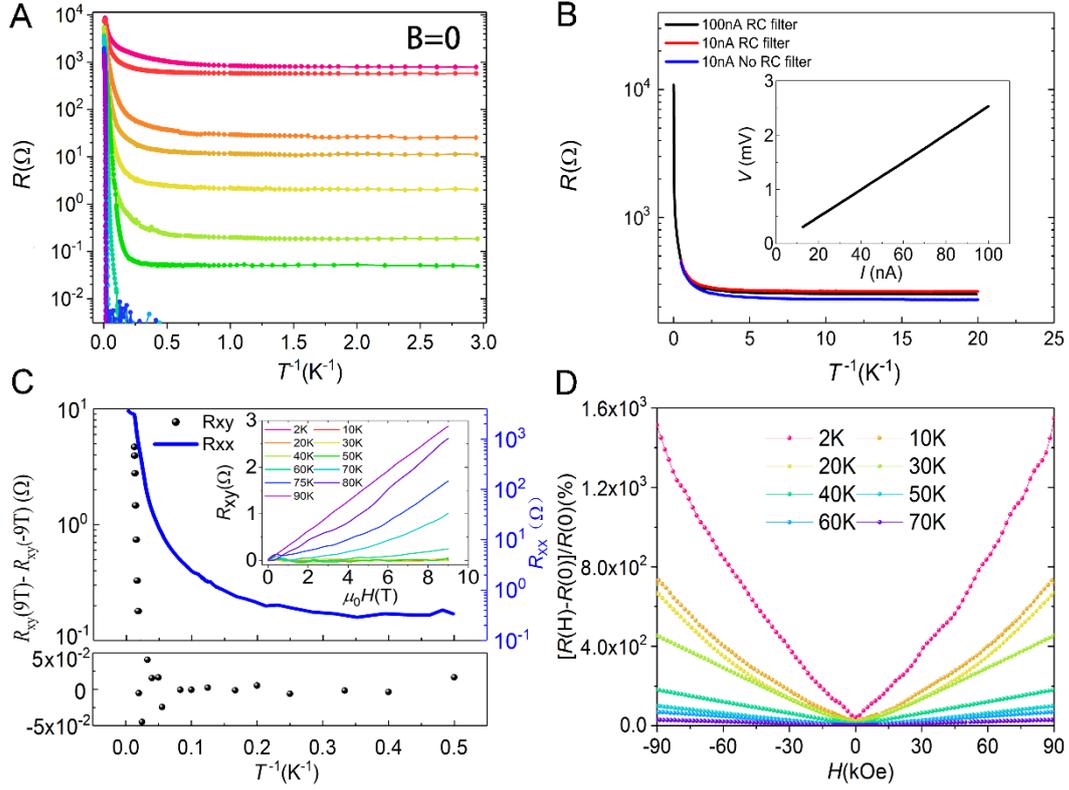

**Fig. 2. Superconductor to anomalous metal phase transition.** (**A**) Arrhenius plot of all the anomalous metallic state films and superconducting films. (**B**) Arrhenius plot of longitudinal resistance ($R_{xx}$) down to 50 mK with and without filters of an anomalous metallic state film. Inset: I-V curve of the anomalous metallic state film. The good linearity below 100 nA indicates the measurements is within ohmic region. (**C**) Arrhenius plot of Hall resistance ($R_{xy}$) and longitudinal resistance ($R_{xx}$) of a represented anomalous metallic state film. The Hall resistance ($R_{xy}$) goes to zero at around 50 K. Inset: Hall resistance ($R_{xy}$) of a represented anomalous metallic state film. (**D**) Giant positive magneto-resistance of a represented anomalous metallic state film.

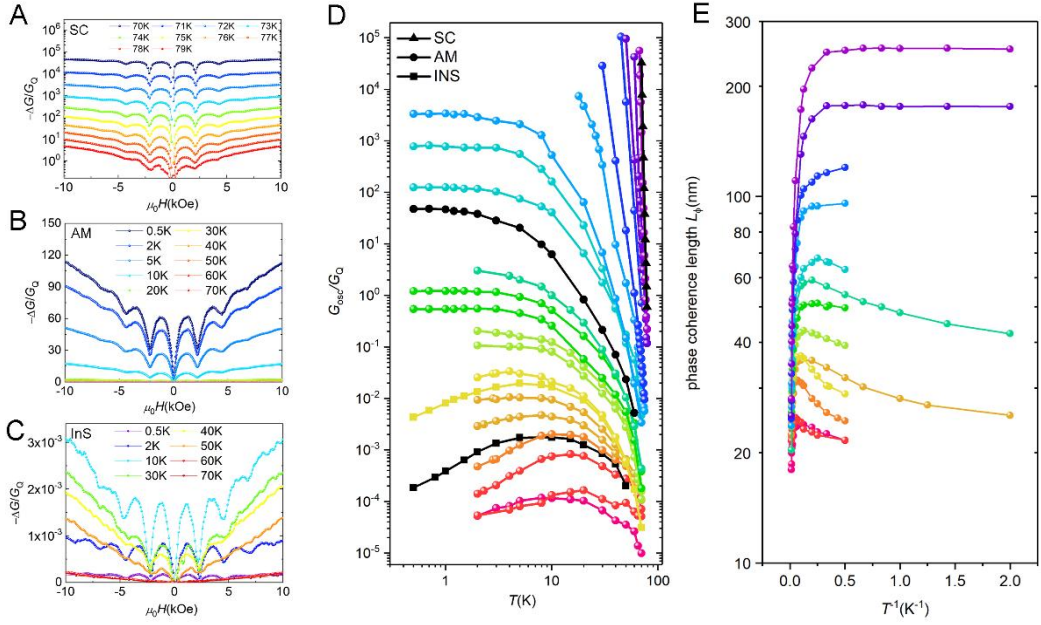

**Fig. 3. The evolution of Cooper pairs coherence through the QPT. A-C,** Perpendicular negative change of magneto-conductance $-\Delta G/G_Q$ of three representative YBCO nanopore films at various temperatures from 70 K down to 0.5 K: (**A**) superconducting state film (SC), (**B**) anomalous metallic sate film(AM), (**C**) insulating state film (INS). Up to four magneto-conductance oscillations are detected with a constant period of $H_0 \sim 2.25$ kOe corresponding to one superconducting flux quantum per unit cell of the nanopattern. (**D**) Temperature dependence of magneto-conductance oscillations amplitude of all the YBCO films in logarithm scale. For anomalous metal state films, $G_{osc}$ saturates with decreasing temperature at around 5 K. By contrast, $G_{osc}$ diverges at low temperatures for superconducting state films, and $G_{osc}$ peaks and drops at low temperatures for insulating state film. Three representative YBCO films are shown in black curve. (**E**) Deduced phase coherence length of insulating and some metallic states films with $G_{osc}$ smaller or comparable to quantum conductance. In the anomalous metallic regime, the phase coherence length $L_\phi$ saturate at lower temperatures. While in the insulating regime, the phase coherence length $L_\phi$ decrease with decreasing temperature.

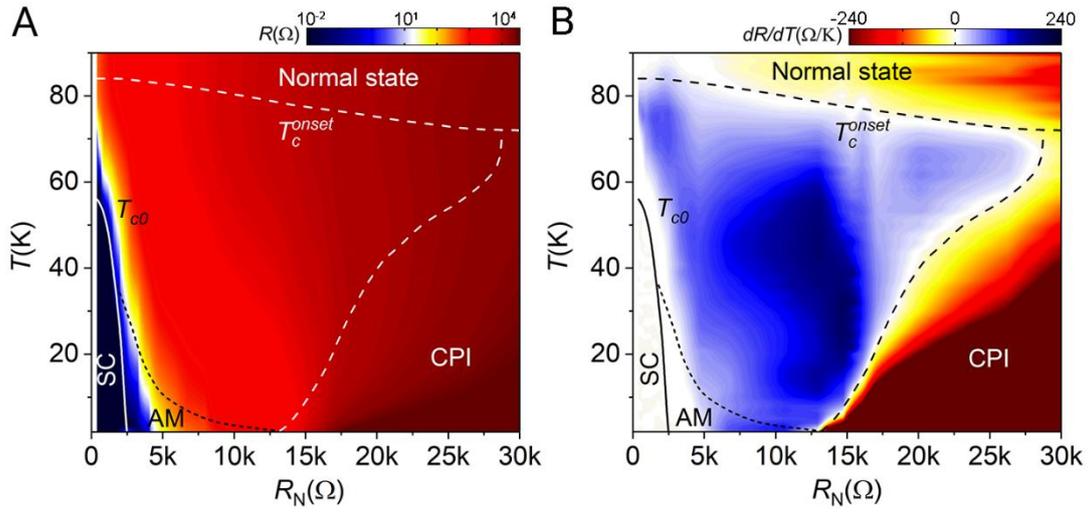

**Fig. 4. Phase diagram of disorder tuned QPT in nanopatterned high-Tc superconductors.**
Colour-scaled map of the resistance (**A**) and differential resistance (**B**) plot as a function of temperature and the normal state resistance. The normal state and other states are separated by $T_c^{onset}$ which is defined by a dashed line. Below $T_c^{onset}$, with increasing the normal resistance, the system undergoes SC phase (zero resistance state within the instrument resolution), AM phase (residual resistance plateau at low temperatures), and CPI phase (Cooper pair localized insulator). The crossover between AM and resistance drop region $(dR_s/dT > 0)$ is separated by a short dash line (ill defined). When approaching zero temperature, with increasing normal resistance, the system evolves from AM to CPI separated by a quantum critical point around $R_N$ ~ 13 kΩ.

Supplementary Materials for

# "Quantum coherence across a superconductor-anomalous metal-insulator quantum phase transition"

Chao Yang[1,*], Yi Liu[2,3,*], Yang Wang[1], Liu Feng[1], Qianmei He[1], Jian Sun[2,3], Yue Tang[2,3], Chunchun Wu[1], Jie Xiong[1†], Wanli Zhang[1], Xi lin[2,3], Hong Yao[3,4,5], Haiwen Liu[6], Gustavo Fernandes[7], Jimmy Xu[7,8], James M. Valles Jr.[8†], Jian Wang[2,3,5,9,10†], and Yanrong Li[1]

*These authors contributed equally to this work.
†To whom correspondence should be addressed. E-mail:
jianwangphysics@pku.edu.cn (J.W.);
james_valles_jr@brown.edu (J.M.V.J);
jiexiong@uestc.edu.cn (J.X.)

[1]State Key Laboratory of Electronic Thin Films and Integrated Devices, University of Electronic Science and Technology of China, Chengdu 610054, China.

[2]International Center for Quantum Materials, School of Physics, Peking University, Beijing 100871, China.

[3]Collaborative Innovation Center of Quantum Matter, Beijing 100871, China.

[4]Institute for Advanced Study, Tsinghua University, Beijing 100084, China.

[5]State Key Laboratory of Low Dimensional Quantum Physics, Tsinghua University, Beijing 100084, China.

[6]Center for Advanced Quantum Studies, Department of Physics, Beijing Normal University, Beijing 100875, China.

[7]School of Engineering, Brown University, 182 Hope Street, Providence, RI02912, USA.

[8]Department of Physics, Brown University, 182 Hope Street, Providence, RI 02912, USA.

[9]CAS Center for Excellence in Topological Quantum Computation, University of Chinese Academy of Sciences, Beijing 100190, China.

[10] Beijing Academy of Quantum Information Sciences, West Bld. #3, No. 10 Xibeiwang East Rd., Haidian District, Beijing 100193, China

This supplement contains:

Materials and Methods

Theory and Calculation

Figs. S1 to S6 and captions

References

**This file is in Word format.**

**Materials and Methods:**

The YBa$_2$Cu$_3$O$_{7-\delta}$ films were grown on (001)-oriented substrates SrTiO$_3$ by magnetron sputtering at T = 700 °C under the pressure of 30 Pa with an argon/oxygen ratio 2:1. The growth rate was very low (~0.8nm/min). Then a 200 nm thick anodized aluminum oxide (AAO) membrane mask was transferred to YBCO thin film. The etching recipe includes 200 W rf power, 1800W inductive coupling plasma (ICP), 40 sccm Ar, 30 sccm CHF$_3$, 20 mTorr (1Torr ≈ 133Pa) pressure. After RIE, hexagonal array of holes morphology of AAO was duplicated to our thin films, which can be tuned from superconducting state to an insulating state by increasing the reactive ion etching time from 20s to 160s (Fig.S6).

AAO templates were prepared in a two-step anodization process. Briefly, polished Al foils were anodized in 0.3 M oxalic acid solution at 40V in 2-6 °C for 5 h. Then the anodic oxide layer was removed in a mixture of H$_3$PO$_4$ (6 wt%) and H$_2$CrO$_4$ (1.8 wt%) at 60 °C. It was anodized again for a short time using the same solution and voltage. Subsequently, a poly methyl methacrylate (PMMA) layer was coated on AAO from a PMMA/methylbenzene solution. The Al layer was removed in mixture of CuSO$_4$ and HCl solutions. The removal of the thin barrier layer was carried out in H$_3$PO$_4$ solution (5 wt%) at 30 °C. PMMA/AAO sheets were put in acetone so that PMMA was dissolved leaving AAO membrane suspending in acetone. The AAO was transferred to YBCO films in the liquid and attached on them with the quick drying of the acetone.

The standard four-probe measurements were carried out in a commercial Physical Property Measurement System (PPMS-16T, Quantum Design) with $^3$He cryostat option for temperature down to 0.34 K. The excitation electrical current as low as possible (5 nA~500 nA) was applied in the ab-plane for the transport measurements while the magnetic field was applied along the c-axis with sweeping rate 0.04 T min$^{-1}$.

The transport experiments extending to extreme low temperature were carried out in a dilution refrigerator (CF-CS81-600; Leiden Cryogenics BV) with a base refrigerator temperature below 10 mK. The thermometry from 1.4 K to 50 mK was based on a commercial thermometer and checked by a cerium magnesium nitrate susceptibility thermometer. Home-made resistor-capacitor (RC) filters and silver-epoxy filters were used to cool down sample electron temperature. Above 22 mK, the electron temperature was equal to the refrigerator temperature confirmed by the temperature dependence of longitudinal resistance in fractional quantum Hall states. The measurements were performed by using standard lock-in techniques, with an alternating current excitation of 10 nA~100 nA at 11 Hz.

**Theory and Calculation:**

To estimate the phase coherence length $L_\phi$, we approximately use the formula $G_{\text{osc}} = C \frac{4e^2}{h} \frac{L_T}{\pi r} \sqrt{\frac{L_\phi}{\pi r}} \exp\left(-\frac{\pi r}{L_\phi}\right)$ for quasi-particle quantum interference (*1*), where $C$ is a geometrical factor ~ 1, $L_T$ is the quasi-particle relaxation length (thermal length), r is half of center-to-center hole spacing ~ 50 nm, $h$ is Planck's constant and $e$ is the electron charge. Since phase coherence time $\tau_\phi^{ee}(T)$ is identical to the quasi-particle relaxation time $\tau_{in}(T)$ (*2*), and $L_T = \sqrt{D\tau_{in}}$, $L_\phi = \sqrt{D\tau_\phi^{ee}}$, we deduce $L_T = L_\phi$. Thus $G_{\text{osc}} = \frac{4e^2}{h} \left(\frac{L_\phi}{\pi r}\right)^{1.5} \exp\left(-\frac{\pi r}{L_\phi}\right)$.

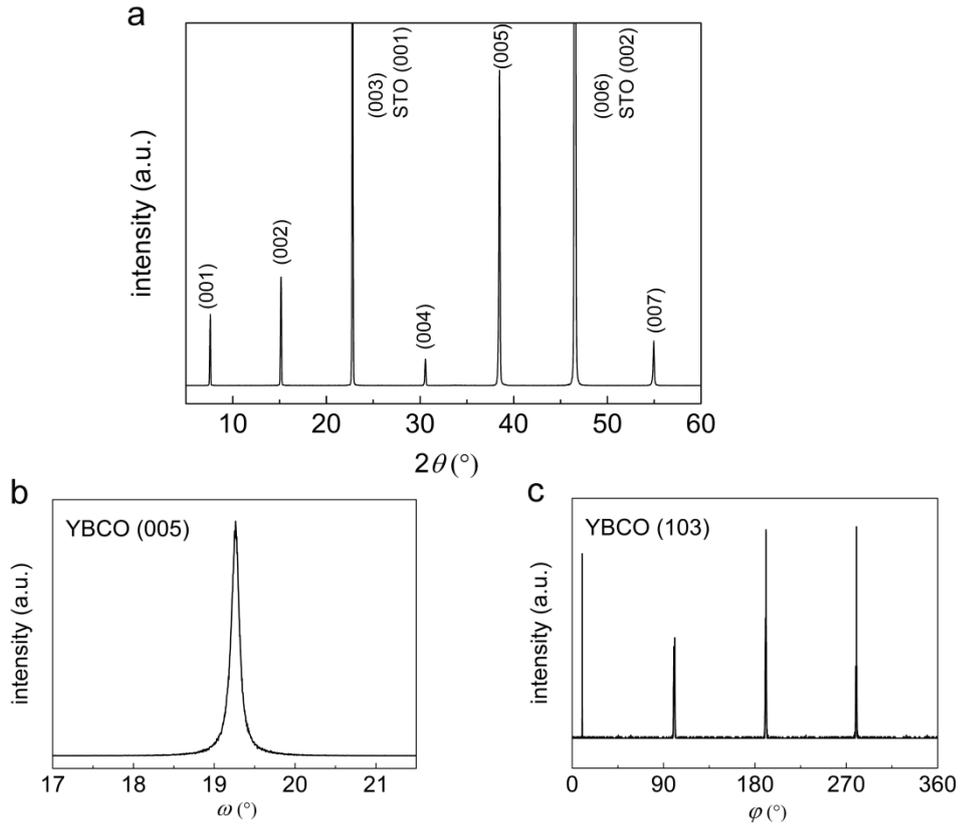

**Fig. S1.** (**A**) X-ray diffraction θ-2θ patterns of YBCO thin films without patterning. (**B**)(**C**) ω-scan and φ-scan of YBCO thin film without patterning. The full width at half-maximum (FWHM) values of out-of-plane and in-plane are around $0.13°$ and $0.9°$, respectively, and four symmetrical peaks in φ-scan reveal a good quality of YBCO thin films.

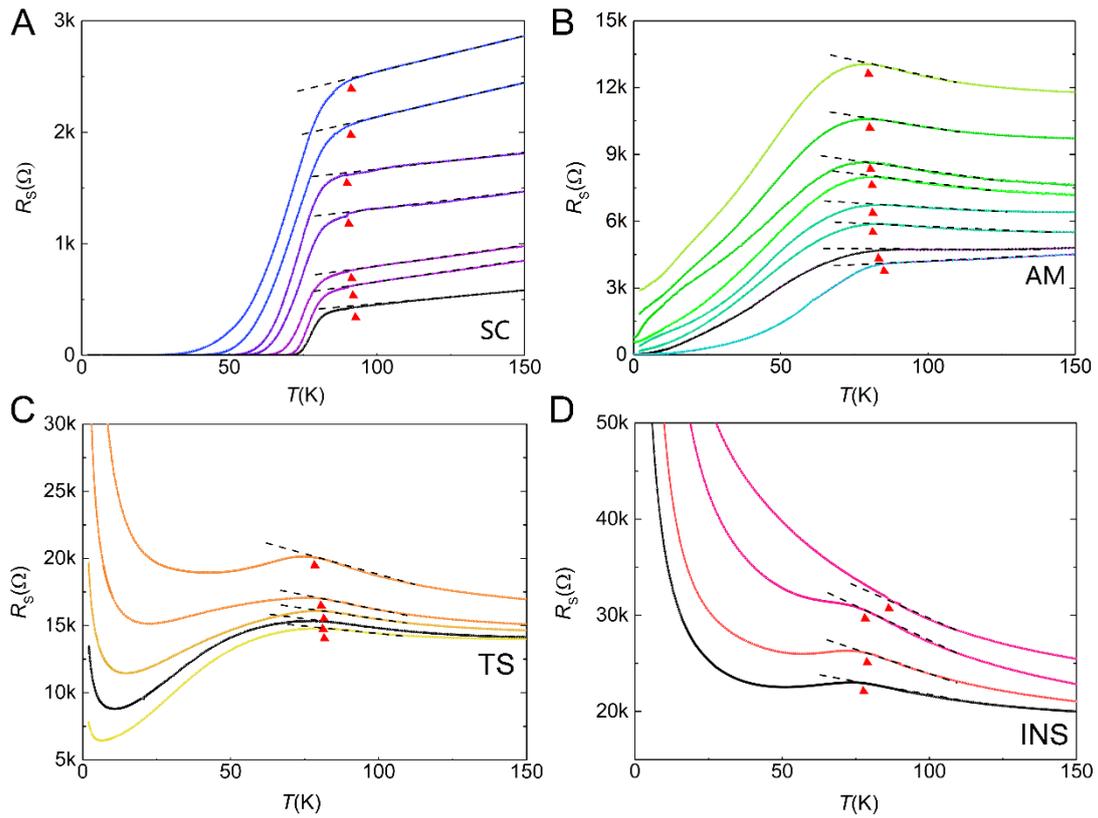

**Fig. S2.** $R_S(T)$ **curves plotted in linear scale for YBCO nanopore films**. (**A**) Superconducting state films (SC), (**B**) Anomalous metallic state films (AM), (**C**) Insulating state films with a resistance drop (Transitional state (TS)), and (**D**) insulating state films (INS). Four representative films as shown in Fig. 1 are labelled as black curves. The superconducting onset critical temperature $T_c^{onset}$ is defined as the temperature when sheet resistance $R_S(T)$ curve deviates from the linear extrapolation of the normal state, marked by a red triangle.

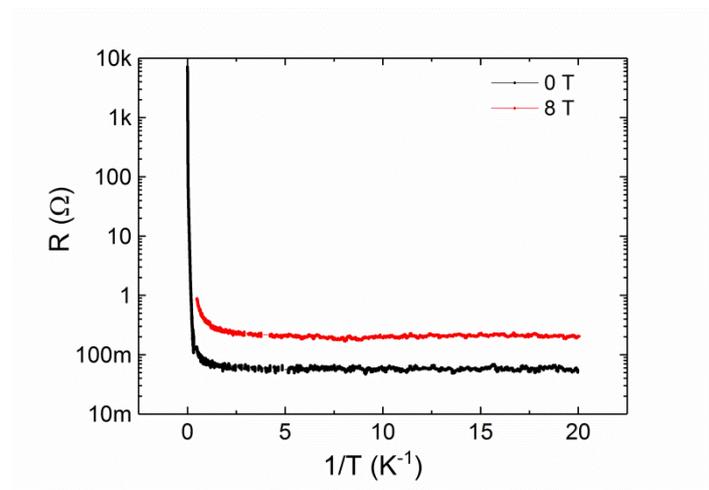

**Fig. S3.** Arrhenius plot of the resistance down to 50 mK of an anomalous metallic state film measured in a well-filtered dilute refrigerator at zero magnetic field and 8T.

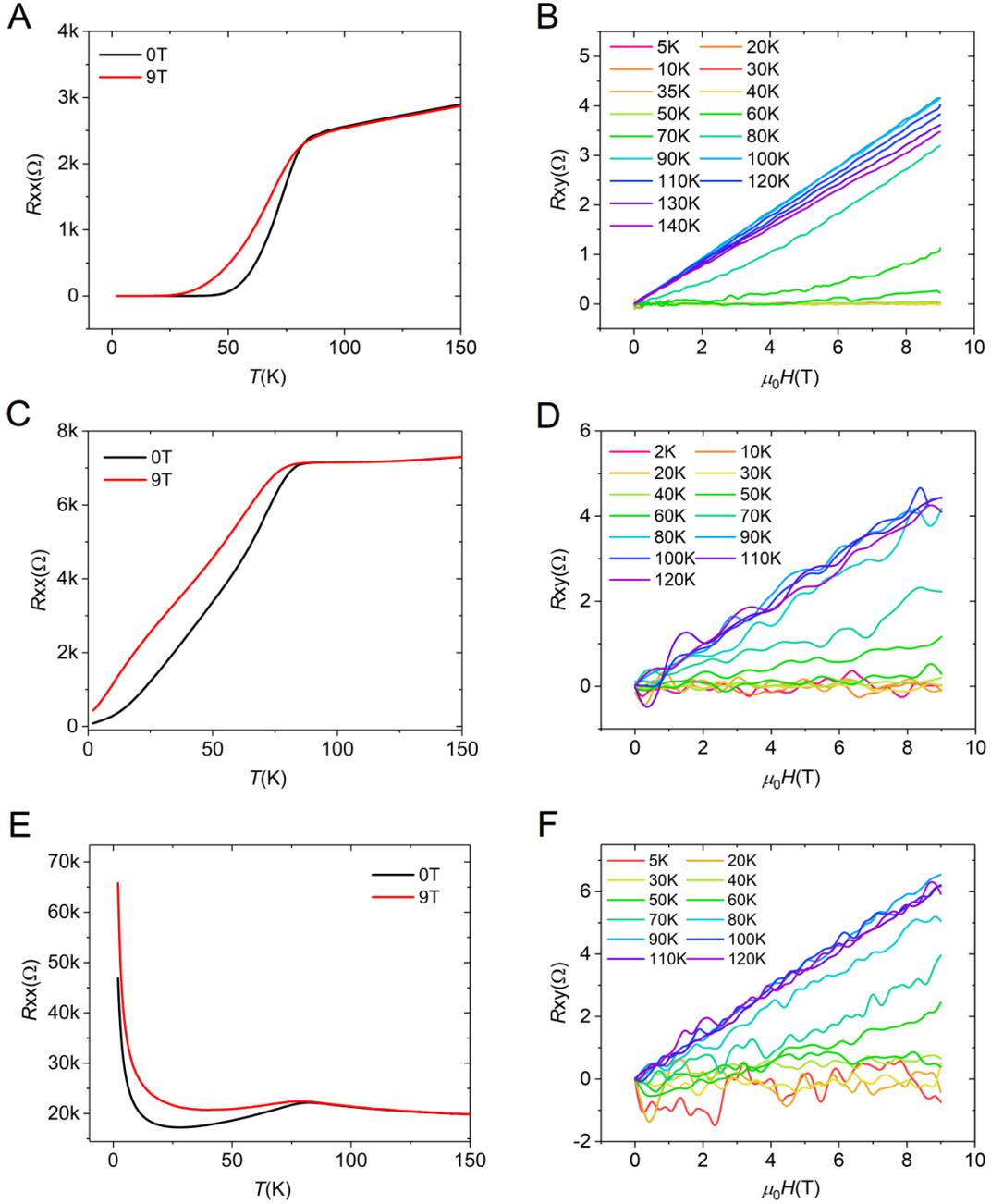

**Fig. S4. The temperature dependence of longitudinal resistance ($R_{xx}$) and Hall resistance ($R_{xy}$) (A)(B)** superconducting state film, **(C)(D)** anomalous metal state film and **(E)(F)** insulating state films. The Hall resistance goes to zero for superconducting state, anomalous metallic state and insulating state within the measurement resolution at around 50K. The absence of Hall resistance at low temperatures reveals the particle–hole symmetry in superconducting state, anomalous metallic state and insulating state, indicating both the anomalous metallic state and insulating state are bosonic, which is consistent with detecting charge 2e quantum oscillations just below $T_c^{onset}$. Interestingly, the magnitude of Hall resistance above $T_c^{onset}$ is almost the same indicating the carrier density remains the same during the QPT.

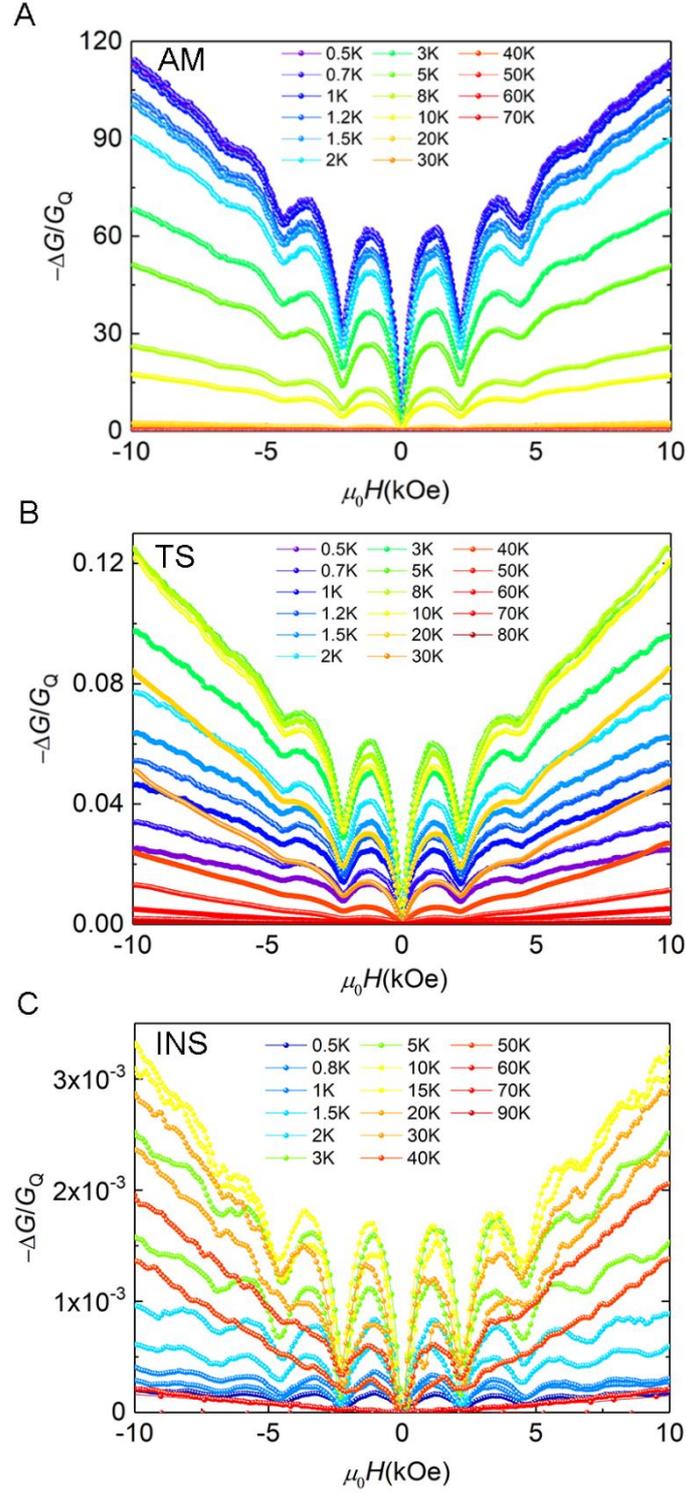

**Fig. S5. Full data of perpendicular magneto-conductance oscillations Δ$G$/$G_Q$ of three representative YBCO films at various temperatures from 90 K down to 0.5 K in the main text.** (**A**) Anomalous metallic state films (AM), (**B**) Transitional state (TS), (**C**) Insulating state films (INS).

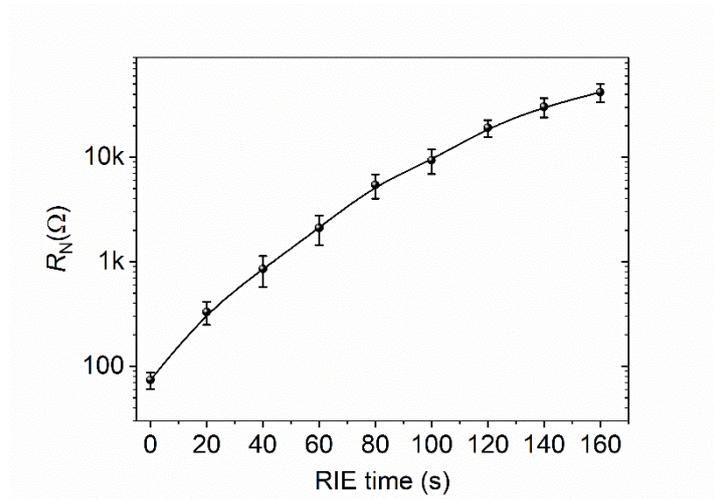

**Fig. S6. Dependence of normal state sheet resistance on reactive ion etching (RIE) time.** After RIE, hexagonal array of holes morphology of AAO was duplicated to YBCO thin films, the normal state resistance increases with increasing etching time from 20 s to 160 s as the systems undergo quantum phase transitions. Error bars represent standard deviations of normal state resistance.

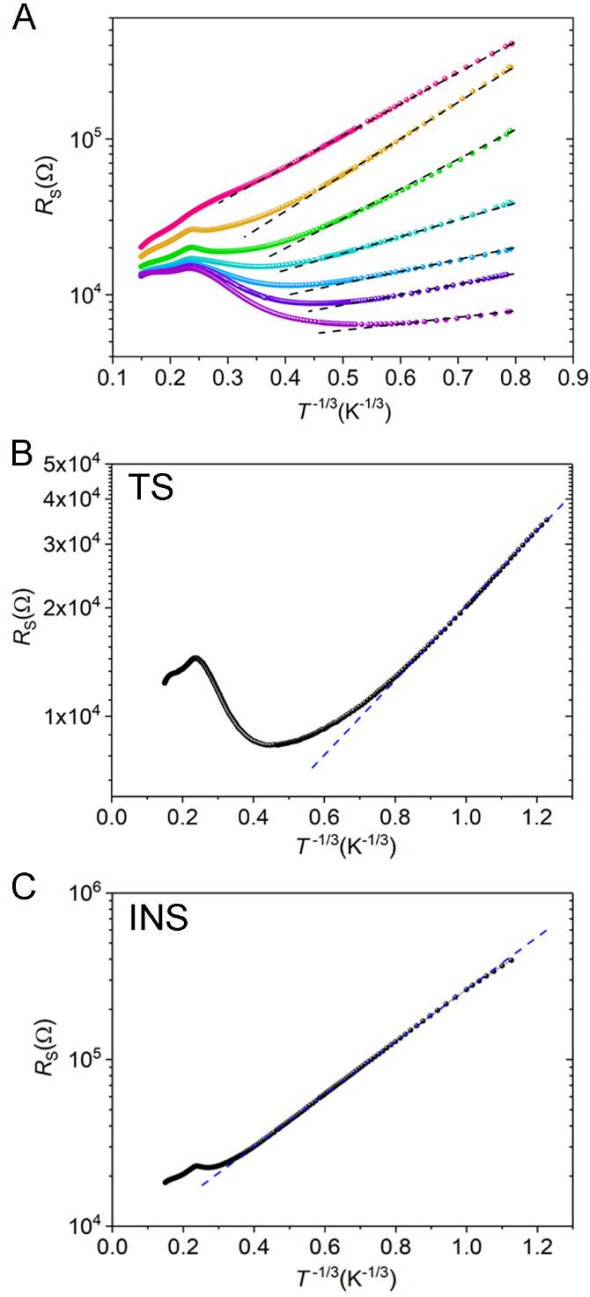

**Fig. S7. Electrical transport property of nanopore YBCO thin films showing resistance upturn at low temperatures.** (**A**), The temperature dependence of insulating upturn of $R_S(T)$ curves meet the two dimensional variable-range hopping (VRH) theory $R_S = R_0 \exp(T_0/T)^{1/3}$. (**B, C**) To further scrutinize this behavior, we extend the transport measurements to lower temperature regime for representative films TS and INS.

## Reference:

1. F. Pierre, *et al.*, *Phys. Rev. B* **68**, 085413 (2003).
2. E. Akkermans, G. Montambaux *Mesoscopic Physics of Electrons and Photons* (Cambridge Univ. Press, Cambridge, 2007).